\documentclass[prb,twocolumn,groupeaddress]{revtex4-2}

\usepackage{amsmath}
\usepackage{graphicx}
\usepackage[colorlinks=true, allcolors=blue]{hyperref}

\begin{document}

\title{Low-energy perspective of interacting electrons in the normal state of superconducting bilayer nickelate}
\author{Frank Lechermann}
\affiliation{Institut f\"ur Theoretische Physik III, Ruhr-Universit\"at Bochum,
  D-44780 Bochum, Germany}
\author{Steffen B\"otzel}
  \affiliation{Institut f\"ur Theoretische Physik III, Ruhr-Universit\"at Bochum,
  D-44780 Bochum, Germany}
\author{Ilya M. Eremin}
\affiliation{Institut f\"ur Theoretische Physik III, Ruhr-Universit\"at Bochum,
  D-44780 Bochum, Germany}

\pacs{}
\begin{abstract}
  Developing a low-energy model is essential for understanding unconventional superconductivity in bilayer
  nickelate La$_3$Ni$_2$O$_7$. Here, we analyze distinct low-energy scenarios of the normal state by downfolding
  the ab-initio determined band structure and applying the mean-field regime of rotational-invariant slave-boson theory.
  We compare models based on the single-site two-orbital, the two-site four-orbital and a proposed minimal cluster (MC)
  picture. The latter builds up on three adapted orbitals located on the sites of the basic Ni-O-Ni
  cluster across the bilayer. Intriguing interplay between the Hund coupling $J_{\rm H}$ and the interlayer
  exchange $J_{\perp}$ is encountered in the multiorbital multi-site problem. While the tendency for interlayer
  Ni-$d_{z^2}$ singlet formation is pronounced, a complete localization remains hindered by the coupling to
  the Ni-$d_{x^2-y^2}$ orbitals. The correlation physics in the MC picture is peculiar with respect to the
  effective interorbital/site exchange.
\end{abstract}

\maketitle

\section{Introduction}
\begin{figure}[t]
      \includegraphics[width=8.5cm]{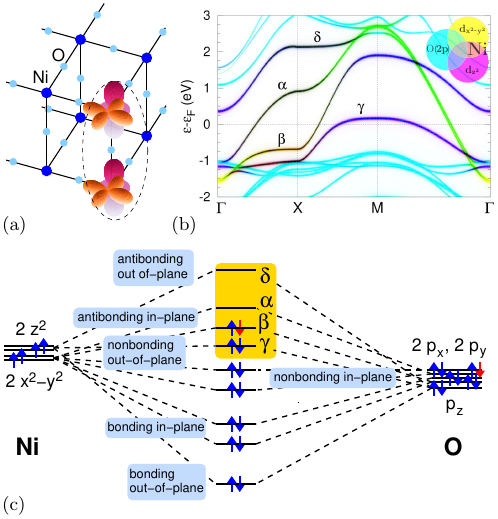}
      \caption{Low-energy states of normal-state La$_3$Ni$_2$O$_7$.
        (a) Sketch of bilayer part with Ni (blue) and O (lightblue) sites.
        Ni-$e_g$ frontier orbitals of $d_{x^2-y^2}$ (orange) and $d_{z^2}$ (pink) type are highlighted. Dashed ellipse marks elementary two-site, two-orbital cluster. (b) DFT bands in fatspec representation for $I4/mmm$ structural
        case~\cite{luhongwang24}. (c) Basic MO energy diagram based on 2 Ni-$e_g^2$, 2 O-$p_{x,y}$ and
        1 O-$p_z$ orbital. Red down spin: ligand hole (understood distributed over whole O$(2p)$ manifold). Golden frame encloses the four states that translate into the $\alpha,\beta,\gamma,\delta$ bands shown in (b).} \label{fig1}
\end{figure}

High-$T_{\rm c}$ superconductivity has been discovered in two regimes of the bilayer
Ruddelsden-Popper nickelate La$_3$Ni$_2$O$_7$. First, under high pressure $p>14$\,GPa~\cite{sun23}
with onset $T_{\rm c}\sim 80$\,K, and very recently also in compressively strained thin films
with onset $T_{\rm c}\sim 40$\,K~\cite{Ko24,zhou-ambient24}. These findings have the potential to inaugarate a
novel area in the research field of unconventional Cooper-pairing. As in akin high-$T_{\rm c}$ cuprate
systems, the bilayer nickelate is subject to strong electronic correlations, dominantly
taking place in the charge-transfer influenced $e_g:\{d_{z^2},d_{x^2-y^2}\}$ subshell of the
transition metal $3d$ manifold. First spectroscopic
characterizations as well as theory assessments~\cite{luohu23,lechermann23,shilenko23,chen-jul23,zhanglin23,Zhang2023second,sakakibara24,christiansson23,yangwangwang23,lupan23,liumei23,yangzhang23,xing-zhou24,jianghuo23,luobiao23,geisler24,oh24,lange24,boetzel24,ryee24,wangjiang24,Savrasov2024,jianghou24,labollita24,chenyangli24,ouyang24,liaowanglei24,wangkunzhang24,nomura25} yet point to obvious differences
between the low-energy electronic structure of the bilayer nickelate and the structurally
similar cuprate compounds. While the latter are apparently dominated by the single Cu-3$d_{x^2-y^2}$
orbital bonding to O$(2p)$, in the former both, Ni-3$d_{x^2-y^2}$ and Ni-3$d_{z^2}$ with their
respective bonding to O$(2p)$ mark the electronic states near the Fermi level (see Fig.~\ref{fig1}a,b).
Thereby, especially the explicit role of the Ni-$d_{z^2}$ degree of freedom for superconductivity
remains ambiguous.
%%%%%%%%%%%%%%%%%%%%%%%%%%%%%%%%%%%%%%%%%%%%%%%%%%%%%%%%%%%%%
%%%%%%%%%%%%%%%%%%%%%%%%%%%%%%%%%%%%%%%%%%%%%%%%%%%%%%%%%%%%%%

To get a better insight it is wise to consider a basic molecular-orbital (MO) picture of superconducting
La$_3$Ni$_2$O$_7$ as drawn from previous first-principles calculations~\cite{lechermann23,shilenko23,chen-jul23}
and shown in Fig.~\ref{fig1}c.
Considering only hoppings between nearest-neighbor Ni and O sites in the primitive cell, restricting
the discussion to the twofold $e_g$ subshell for Ni$(3d)$, and starting from the canonical Ni$^{2+}$
and O$^{2-}$ ionic oxidation states, one arrives at the following scenario: the equivalent
Ni-$d_{x^2-y^2}$ orbitals in the upper/lower part of the bilayer hybridize with their neighboring
O-$p_{x,y}$, respectively, accomodating in total ten electrons in six in-plane MOs. These include two bonding,
two O-dominated nonbonding and two antibonding orbitals. Note that finally, each resulting twofold MO
sector bilayer splits into a symmetric and antisymmetric state. On the other hand, the Ni-$d_{z^2}$
orbitals in the upper/lower part of the bilayer both hybridize with the bridging O-$p_z$ orbital,
leading to three additional out-of-plane MOs for four electrons: strongly-split bonding and antibonding
states, as well as a Ni-dominated nonbonding state inbetween. As a result, all the in-plane
bonding/nonbonding MOs, as well as the deepest-lying out-of-plane bonding MO, are completely filled.
This leaves four frontier orbitals close to the Fermi level for four remaining electrons.  
However, through charge neutrality from a canonical La$^{3+}$ oxidation state, the described key
Ni$_3$O$_5$ subsystem in the unit cell of La$_3$Ni$_2$O$_7$ hosts only 13 instead of 14 valence electrons.
As confirmed by recent experiments~\cite{dong24},
this missing electron formally acts as a ligand-hole among O$(2p)$. Thus there remain only three electrons
in the four frontier MOs, {\sl aka} in the four resulting low-energy bands. The lower edge of this band
manifold is then formed by the nearly-filled $d_{z^2}$-dominated non-bonding (and therefore well-flattened)
$\gamma$ band, while the upper edge consists of the antibonding $d_{z^2}$-$p_z$ $\delta$ band. Enclosed
inbetween are the two antibonding $d_{x^2-y^2}$-p$_{x,y}$ bilayer-split $\alpha,\beta$ bands.

The main goal of our manuscript is to provide further insights on the the bilayer La$_3$Ni$_2$O$_7$ electronic
states and their correlations based on the realistic low-energy dispersions. In particular, we study the
correlation effects depending on Hubbard $U$, Hund exchange $J_{\rm H}$, filling $n$, as well as possible
interlayer self-energies using a state-of-the-art slave boson approach. We identify
key competing effects, which lead to an inherent frustration characteristics. We compare the results for
already existing model Hamiltonians for this system, and dervive an alternative truly low-energy cluster
model. It is argued that this minimal-cluster model captures the very essentials of the challenging
bilayer nickelate compound prone to superconductivity.

\section{Theoretical Approach}
\subsection{Basic low-energy Hamiltonian}
In order to perform a realistic low-energy investigation, the DFT band structure is
downfolded to the introduced four $\alpha,\beta,\gamma,\delta$ bands by a maximally-localized
Wannier construction~\cite{marzari12}. A four-band tight-binding (TB) Hamiltonian results from this
procedure. We thereby rely on the experimental
orthorhombic $Fmmm$ structure for pressurized La$_3$Ni$_2$O$_7$~\cite{sun23}, as already in the
previous work Ref.~\onlinecite{lechermann23}. Yet especially the interlayer separation between
the bilayers slightly varies between different structural refinements, and also, recent experimental
data points to a tetragonal $I4/mmm$ symmetry. The latter symmetry is also assumed in the recent
thin-film experiments~\cite{Ko24,zhou-ambient24}. For specific cases we therefore compare the results
to those obtained for an $I4/mmm$-symmetric TB Hamiltonian. For concreteness, that one is
derived by using the experimental crystal structure from Ref.~\onlinecite{luhongwang24} at 24.6\,GPa
and by subsequent ab-initio optimization of the atomic positions. The underlying DFT calculations are
performed with a mixed-basis pseudopotential code~\cite{elsaesser90,lechermann02,mbpp_code}, using
norm-conserving pseudopotentials and a basis consisting of plane waves as well as modified atomic
functions for La$(5d)$, Ni$(3d)$ and O$(2s,2p)$. Scalar-relativistic effects are taken care
of by frozen-core pseudopotentials, but the effect of spin-orbit coupling is neglected.

It is important to realize from the beginning, that the Wannier basis for the obtained
basic TB Hamiltonian is surely different from an atomic-orbital basis. While the four Wannier
orbitals are still centered on two symmetry-equivalent Ni basis sites in the unit cell
(one site for each layer),
they have additional strong weight from nearby O$(2p)$. Therefore they neither represent the
atomic Ni states nor the MO states, both utilized in Fig.~\ref{fig1}c. Correspondingly, their occupation 
also does not resemble those fillings. For instance, while each of the {\sl atomic} Ni-$e_g$
orbitals is very close to half-filling (see left part of Fig.~\ref{fig1}c), this will not
be the case for the Wannier orbitals. The latter individual occupations will sum up to three
electrons, as true for the bands emerging from the MO picture. For reasonable wording, we will
in the following still name the Wannier orbitals as $d_{z^2}$ and $d_{x^2-y^2}$, but the
aforementioned differentiation has to be kept in mind. 

The interacting low-energy Hamiltonian may be cast in the following form
\begin{equation}
H_{\rm basic}
= H^{\hfill}_{\rm kin} + \sum_u\left( H^{(u)}_{\rm int} + H^{(u)}_{\rm orb}\right)\;,
\label{eq:ham1}
\end{equation}
using the label $u$ for the unit cell. It incorporates the kinetic part $H_{\rm kin}$, 
the local-interacting part $H_{\rm int}$ and a further non-interacting local-orbital
contribution $H_{\rm orb}$. The kinetic bilayer Hamiltonian may be written with
hoppings $t$ as
\begin{equation}
  H^{\hfill}_{\rm kin} = \sum_{ij,\ell\ell',mm',\sigma}
  \hspace*{-0.55cm}^\prime\hspace*{0.3cm}t^{\hfill}_{ij,\ell\ell',mm'}
\,c^\dagger_{i\ell m\sigma} c^{\hfill}_{j\ell' m'\sigma}\;,
\label{eq:hamkin}
\end{equation}
where $i,j$ are lattice sites within a layer $\ell,\ell'=1,2$ of the bilayer, 
$m,m'=d_{z^2},d_{x^2-y^2}$ and $\sigma=\uparrow,\downarrow$. Notably, the
prime at the sum sign renders clear that this kinetic form is liberated
from all intra-unit-cell contributions.
For $H_{\rm int}$ we choose the Slater-Kanamori form to describe local
interactions among the $e_g$ orbitals within a layer, respectively, i.e.
\begin{eqnarray}      
H^{(u)}_{\rm int}&=&U\sum_{\ell,m} n_{\ell m\uparrow}n_{\ell m\downarrow}    
+\sum_{\ell,m\neq m',\sigma}              
\left\{\frac{U'}{2} \, n_{\ell m\sigma} n_{\ell m' \bar \sigma}\right.\nonumber\\     
&&+\, \frac{U''}{2} \,n_{\ell m\sigma}n_{\ell m' \sigma}+           
J_{\rm H}\, c^\dagger_{\ell m \sigma} c^\dagger_{\ell m' \bar\sigma} 
c^{\hfill}_{\ell m \bar \sigma} c^{\hfill}_{\ell m' \sigma}\nonumber\\    
&&+\left. J_{\rm H}\,c^\dagger_{\ell m \sigma} c^\dagger_{\ell m \bar \sigma}  
 c^{\hfill}_{\ell m' \bar \sigma} c^{\hfill}_{\ell m' \sigma}\right\}\;,
\label{eq:intham}          
\end{eqnarray} 
with $U'=U-2J_{\rm H}$ and $U''=U-3J_{\rm H}$. The remaining Hamiltonian part
$H_{\rm orb}$ carries all the intra-unit-cell terms left out in
eq.~(\ref{eq:hamkin}).

\subsection{Rotationally invariant slave bosons}
The interacting Hamiltonian is solved by the rotationally invariant
slave-boson (RISB) scheme~\cite{li89,lec07,bunemann07,lan17,pie18,fac18} on the mean-field 
level. In essence, this many-body technique builds up on a fragmentation of the
quasiparticle (QP) character and the local-excitation character on the operator level,
here formally reading
\begin{equation}
c_{\ell m\sigma}^\dagger=R(\{\phi\})\,f_{\ell m\sigma}^\dagger\qquad.
\end{equation}
The $f$ degree of freedom describes the fermionic QP state, and the bosonic set $\{\phi\}$
provides access to the local multiplets. In the generic RISB approach, there is one
$\phi$ boson for each pair of local Slater determinants
$(|\{n^{(p)}_{\ell m\sigma}\}\rangle,|\{n^{(p')}_{\ell' m'\sigma'}\}\rangle)=:
(|{\cal A}_p\rangle,|{\cal Q}_{p'}\rangle)$, with $p,p'$ denoting particle sectors. But as we
here focus on normal-state properties, there are only bosons connecting states in
identical particle sectors $p=p'$ considered. The extension to the explicit
superconducting phase is discussed in Ref.~\onlinecite{isi09}. Note that formally, the
index ${\cal A}$ labels a local atomic state and ${\cal Q}$ the QP degree of
freedom~\cite{lec07}.

The RISB electronic self-energy $\Sigma$ is local and consists of a term linear in frequency
as well as a static part. While the former part gives rise to Fermi-liquid properties
of the correlated state, the latter describes the renormalization of local
hybridizations and crystal fields. There are no frequency-dependent Hubbard bands
in the mean-field RISB scheme, however information about the local excitations (e.g.
average multiplet occupations, spin correlations, etc.) within a correlated metal
is still included by the frequency-independent bosonic amplitudes. The method hence
lacks the full-frequency dependence of the dynamical mean-field theory (DMFT) self-energy,
but is still well suited (here at formal $T=0$) for a large class of correlated materials
problems. For instance, it naturally may describe local-moment formation from
$\langle S^2\rangle>0$, whereas e.g. conventional DFT(+U) schemes can only access static
moments with $\langle S\rangle\neq 0$. For further methodological details
see e.g. Refs.~\onlinecite{lec07,pie18}.

Because of its general representation of the local Hilbert space, similar as DMFT the
RISB methodology is directly applicable to cluster formulations of the interacting-electron problem on a lattice~\cite{lec07,ferrero09}. This means the local entity in the approach is not restricted to
a single site, but can easily consist of a cluster of several sites. This feature
is especially useful in the present case of La$_3$Ni$_2$O$_7$, which exhibits
a natural cluster building block on the given lattice (see Fig.~\ref{fig1}a).
The straight three-site cluster of Ni($\ell=1$)-O-Ni($\ell=2$) across the bilayer
translates into an effective two-site-four-orbital cluster when integrating out
the non-Ni-$e_g$ degrees of freedom for the canonical Wannier Hamiltonian. Due to
the isolated bilayer in the Ruddlesden-Popper structure, the effect of
translational-invariance breaking in a cellular-cluster scheme is not an issue.
A cluster-DMFT study of this nickelate bilayer problem has already been performed by
Ryee {\sl et al.}~\cite{ryee24}.
For a cluster-RISB treatment of variational kind of that size, the scaling of the bosonic
Hilbert space becomes demanding. Whereas the number of bosonic degrees of freedom in the
single-site treatment of the two-orbital problem (without using further symmetry constraints)
from summing
over particle sectors $p=0,1,2,3,4$ amounts to $2\times(1+4^2+6^2+4^2+1)=140$, there are
25740 bosonic degrees of freedom in the two-site cluster description with four orbitals
and eight particle sectors. The factor '2' stems from the fact that it proves generally
useful for numerical reasons to treat the condensed $\phi$ boson as a complex number,
hence the original number of degrees of freedom is duplicated.
The exponential Hilbert-space scaling asks for an efficient handling of
the important mixing between RISB variational steps. Whereas for the lower-orbital 
calculations the modified-Broyden method~\cite{van84} is put into practise, a numerically
less costly scheme is needed for four-orbital calculations. In that case, we employ the recursive
implementation of the modified-Broyden method and the direct inversion
in the iterative subspace (DIIS) method~\cite{kawata98}.

\section{Results}
The low-energy interacting picture emerging from the coupled single-site correlation treatment of
$H_{\rm basic}$ is discussed in section~\ref{res1}. In order to account for interlayer self-energies
and also possibly pronounced orbital-selective features in alternative orbital-basis representations,
the two-site cluster results are discussed in section~\ref{res2}. Finally, in
section~\ref{res3} we present a possible minimal-cluster modeling based on only three effective
Wannier orbitals. If not otherwise stated, the data in sections~\ref{res1},\ref{res2} is obtained for
the $Fmmm$ symmetry, and for section~\ref{res3} the $I4/mmm$ symmetry used. All calculations are
performed for a paramagnetic regime.

\subsection{Coupled single-site two-orbital picture\label{res1}}
%%%%%%%%%%%%%%%%%%%%%%%%%%%%%%%%%%%%%%%%%%%%%%%%%%%%%%%%%%%%%%
\begin{figure*}[t]
      \includegraphics[width=\linewidth]{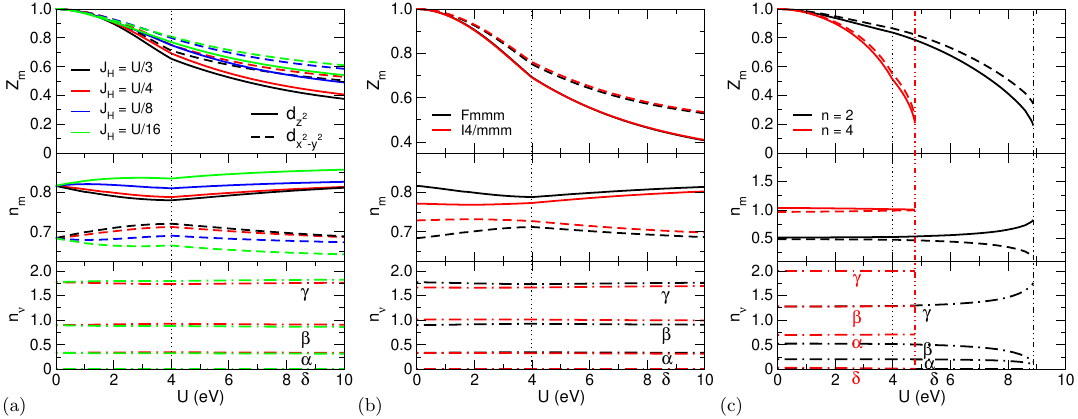}
      \caption{Coupled single-site results for QP weight $Z_m$ (top panels) and orbital occupation
        $n_m$ per site (mid panels) for $d_{z^2}$ (full lines) and $d_{x^2-y^2}$ (dashed lines),
        as well as QP band occupations $n_\nu$ for $\nu=\alpha,\beta,\gamma,\delta$ (see Fig.\ref{fig1}b)
        (bottom panels), respectively. Vertical dotted lines mark onset of fixed
        $J_{\rm H}=4/(3,4,8,16)$\,eV regime for $U>4$\,eV. 
        (a) Dependence on $J_{\rm H}$ for $n=3$. (b) Performance comparison between $Fmmm$ and $I4/mmm$
        low-energy Hamiltonian for $n=3$. (c) Cases $n=2$ and $n=4$, with vertical
        dashed-double-dotted lines denoting respective critical $U$.
        Results in (b,c) are for fixed $J_{\rm H}=U/4$.}\label{fig2}
\end{figure*}	
%%%%%%%%%%%%%%%%%%%%%%%%%%%%%%%%%%%%%%%%%%%%%%%%%%%%%%%%%%%%%
From the orbital on-site levels $\varepsilon_m$, the DFT crystal-field splitting
$\Delta_{\rm CF}=\varepsilon_{z^2}-\varepsilon_{x^2-y^2}$ in the Wannier Hamiltonian
$H_{\rm basic}$ (eq.~(\ref{eq:ham1})) reads $\Delta_{\rm CF}=-161(-76)$\,meV for
the $Fmmm$($I4/mmm$) structural case. Hence the $d_{z^2}$ orbital is lower in energy, and about twice as
much so for the orthorhombic structure compared to the tetragonal one. The interlayer hoppings $t_\perp^m$
are further key energy scales of the problem and amount to $t_\perp^{z^2}=-783(-725)$\,meV and
$t_\perp^{x^2-y^2}=-8(20)$\,meV for $Fmmm$($I4/mmm$). Nonsurprisingly, the $d_{z^2}$($d_{x^2-y^2}$)
orbitals are strongly(weakly) coupled across the bilayer.

In a straightforward step, the treatment of electronic correlations in La$_3$Ni$_2$O$_7$ based on
the four-band model consists in employing the symmetry equivalence of
both Ni ions in the upper and lower layer of the bilayer structure, and performing a multi-single-site
RISB calculation. Thereby, the electronic self-energy is of two-orbital $\{d_{z^2},d_{x^2-y^2}\}$ nature,
but purely Ni on-site.

Before discussing the results as summarized in Fig.~\ref{fig2}, let us first allude to the fact that
the DFT-revealed odd band occupation of $3/8$ for the given four-band manifold is quite peculiar. Since it
amounts to three electrons distributed over two Ni sites, an interaction-driven localization within
a single-site Mott scenario is impossible without site-symmetry breaking. Since such a symmetry breaking
is apparently absent in experiment~\cite{sun23}, we stick to the symmetry-equivalent treatment for the
Ni sites and therefore do not expect a strongly-renormalized QP weight $Z_m=\left[1-\frac{\partial}{\partial\omega}\Sigma_m\,\right]_{\omega=0}^{-1}$,
which would resemble a single-site Mott-critical regime. While very strong correlation effects have been revealed
in the recent advanced-DMFT study~\cite{lechermann23}, those rely dominantly on flat-band driven non-Fermi
liquid physics as induced by the $\gamma$ band and described by the full-frequency dependence of
$\Sigma(\omega)$. Yet such physics is not included in the mean-field RISB approach, which restricts the
description to a Fermi-liquid QP picture of electronic correlations. However, orbital-selective physics is surely
covered by this and one may wonder about a sole Mott localization of $d_{z^2}$. Yet since $t_\perp^{z^2}$
is by far the largest effective Ni-Ni hopping in the system, such an orbital-selective Mott regime is
not exptected in the single-site picture. But in principle, this might be possible in a cluster
treatment of correlations via interl-layer $d_{z^2}$ singlet formation. This will be discussed in
section~\ref{res2}.
%%%%%%%%%%%%%%%%%%%%%%%%%%%%%%%%%%%%%%%%%%%%%%%%%%%%%%%%%%%%%%
\begin{figure}[b]
  \includegraphics[width=\linewidth]{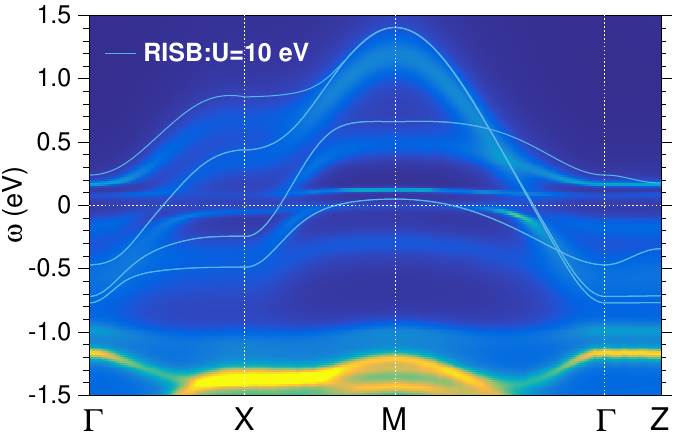}
  \caption{Comparison of the QP bands within RISB for $U=10$\,eV, $J_{\rm H}=1$\,eV (full lightblue lines)
    with the charge-selfconsistent DFT+sicDMFT spectral function~\cite{lechermann23} for the high-pressure
  $Fmmm$ structure of La$_3$Ni$_2$O$_7$.}\label{fig3}
\end{figure}	
%%%%%%%%%%%%%%%%%%%%%%%%%%%%%%%%%%%%%%%%%%%%%%%%%%%%%%%%%%%%%

Figure~\ref{fig2}a displays the main RISB results for the adequate $n=3$ filling of three electrons in
the four bands, depending on the ratio $U/J_{\rm H}$. As expected, the strong-coupling QP weight
$Z_m\sim 0.4-0.6$ remains sizable and correlation-induced renormalization is thus moderate for $U$ up to
10\,eV. A lowering of the Hund exchange $J_{\rm H}$ further weakens the effect of correlations. This is
understood from general Hund-metal physics away from local half filling (see e.g.
Ref.~\onlinecite{georges13} for a review). In general, the $d_{z^2}$ orbital is still stronger correlated
than the $d_{x^2-y^2}$ one. The orbital filling $n_m$ is also larger for $d_{z^2}$ thanks to the crystal-field
splitting (see mid panel of Fig.~\ref{fig2}a). The orbital polarization $n_{z^2}-n_{x^2-y^2}$ increases with
lowering $J_{\rm H}$, since the Hund exchange fosters orbital balancing. This polarization is still
growing for $U>4$\,eV after the physically-resonable fixing of $J_{\rm H}$ to its value at $U=4$. From the
bottom panel of Fig.~\ref{fig2}a, band occupations hardly change with local Coulomb interactions,
but qualitatively, a very minor filling increase of the $\gamma$ band at large $U$ may be stated.

The comparison of those observables for the two structural cases, i.e. $Fmmm$ and $I4/mmm$, as shown
in Fig.~\ref{fig2}b shows no qualitative differences, and the quantitative ones are small, especially
for the QP weights $Z_m$. The larger orbital differentiation from $n_m$ is seed already by the larger
$\Delta_{\rm CF}$, translating also to a slightly larger $\gamma$ band occupation. Thus in general,
for the present level of investigation in this work we do not expect qualitative physics differences
between the orthorhombic and tetragonal bilayer symmetry.

Figure~\ref{fig3} exhibits moreover a comparison of the $n=3$ QP bands as obtained for $U=10$\,eV and
$J_{\rm H}=1$\,eV, with the corresponding charge-selfconsistent DFT+sicDMFT spectral function for
the high-pressure $Fmmm$ phase from Ref.~\onlinecite{lechermann23}. The QP bands are in principle in the
same energy window as the more elaborate low-energy data, pointing indeed to a general renormalization
by electronic correlations of similar order. The stronger correlations effects in the advanced DMFT data,
especially close to the Fermi level and/or for the $\gamma$ band, should originate from the flat-band
(non-Fermi-liquid) physics impact, which is beyond the capabilities of mean-field RISB. Importantly
however, the flat $\gamma$ band is in either case only slightly doped, but still tied to the Fermi level.

Albeit hardly relevant for the concrete experimental system, it proves finally instructive from a model
perspective to study also the nearby integer-filling regimes $n=2$ and $n=4$. Since these quarter- and
half-filled scenarios resemble even electron occupation on the two Ni sites in the unit cell, a standard
single-site Mott transition with increasing $U$ becomes possible. The given observables $Z_m$, $n_m$ and $n_\nu$
as shown in Fig.~\ref{fig2}c indeed report a Mott transition in both cases. For $n=2$, a single electron
on each Ni becomes localized at $U_{\rm c}^{\rm (q)}\sim 8.8$\,eV, notably with strong orbital polarization
towards $d_{z^2}$. At half filling $n=4$, two localized electrons on each Ni site (and hence in each
layer of the bilayer) are formed at $U_{\rm c}^{\rm (h)}\sim 4.8$\,eV. For sure, the band occupations are
also highly different from $n=3$, the $\gamma$ band being always completely occupied at half filling and
on the verge to full occupation for quarter filling. Note that the so far spectator band $\delta$ also becomes
slightly occupied at half filling.

\subsection{Two-site four-orbital cluster picture\label{res2}}
%%%%%%%%%%%%%%%%%%%%%%%%%%%%%%%%%%%%%%%%%%%%%%%%%%%%%%%%%%%%%%
\begin{figure*}[t]
      \includegraphics[width=\linewidth]{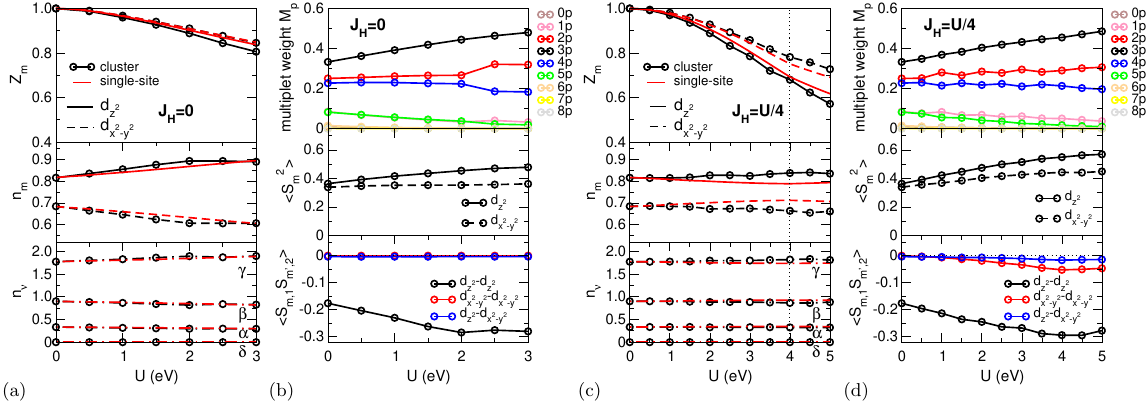}
      \caption{Two-site cluster results for cases $J_{\rm H}=0$ (a,b) and $J_{\rm H}=U/4$ (c,d).
        (a,c) QP weight $Z_m$ (top panels) and orbital occupation
        $n_m$ per site (mid panels) for $d_{z^2}$ (full lines) and $d_{x^2-y^2}$ (dashed lines),
        as well as QP band occupations $n_\nu$ for $\nu=\alpha,\beta,\gamma,\delta$ (see Fig.\ref{fig1}b)
        (bottom panels), respectively. Vertical dotted line in (c) marks onset of the fixed
        $J_{\rm H}=1$\,eV regime for $U>4$\,eV. (b,d) Analysis of the local bosonic amplitudes
        $\{\phi\}$, giving rise to the sum of the multiplet weight $M_p$ among the 0-8 particle
        sectors (top panel) as well as the orbital-resolved local spin moment $\langle S_m^2\rangle$
        (mid panel) and interlayer spin correlation $\langle S_{m,\ell=1}S_{m',\ell=2}\rangle$ (bottom
        panel).}\label{fig4}
\end{figure*}	
%%%%%%%%%%%%%%%%%%%%%%%%%%%%%%%%%%%%%%%%%%%%%%%%%%%%%%%%%%%%%
In order to investigate the role of interlayer correlations in the relevant $n=3$ case, the two-site
cellular-cluster scheme using two orbitals per site within RISB is employed. Note that the original
Wannier-orbital basis is kept as it is, and the interacting Hamiltonian consists of a block-diagonal
Slater-Kanamori form encircling both Ni sites. There are no intersite Coulomb interactions, but
site-offdiagonal self-energy terms $\Sigma_{\ell=1,\ell=2}$ become now possible. 

On the model-Hamiltonian level, the more basic problem of a single-orbital bilayer Hubbard model at half filling
has already been studied quite extensively (e.g. Refs.~\onlinecite{Bulut92,kancharla07,schuwalow12,Lee14,Golor14}).
Albeit there are still open question about aspects of the corresponding phase diagram~\cite{Mou22}, the
main competitors are identified. For $U=0$, the system shows a transition from a metal to a band insulator,
driven by the interlayer hopping $t_\perp$. For reasonably large $U$, a Mott-insulator to band-insulator
transition takes place, with possibly an intermediate metallic phase, when increasing again $t_\perp$.

The present problem of an effective two-orbital bilayer Hubbard model away from half filling is even more
complicated. The interplay of orbital-dependent hoppings with Hubbard $U$ and Hund $J_{\rm H}$ for the
given $3/8$ filling should be intriguing. First, the present $t_\perp$ is quite large for $d_{z^2}$, and
weak for $d_{x^2-y^2}$. Second, the role of $J_{\rm H}$ is at least twofold: it promotes a {\sl common}
local-spin formation in the multiorbital on-site setting and fosters local orbital-filling balancing.
Hence while a large $t_\perp^{z^2}$ is a driving force for interlayer $d_{z^2}$ singlet formation,
Hund exchange balances the orbital-spin response. A straightforward ``chained to $d_{z^2}$'' behavior
of $d_{x^2-y^2}$ is not to be expected. Because, if a $d_{z^2}$ singlet would indeed be realized, for the
present filling scenario, there is still one Hund-frustrated $d_{x^2-y^2}$ electron delocalized within
the bilayer.
And finally, while Fig.~\ref{fig1}b suggests the $d_{z^2}$ subsystem already close to the band-insulating
regime from inspection of the $\gamma,\delta$ bands, the situation is of course much more complex.
There is additional hybridization with $d_{x^2-y^2}$ and most importantly, the $\gamma$ band
is nonbonding with reference to the briding O-$p_z$ orbital (the true antibonding band from this
three-orbital subsystem is deep down in energy in the occupied part of the spectrum). As stated before,
the present $\{d_{z^2}, d_{x^2-y^2}\}$ orbitals are Wannier orbitals, including amount of oxygen content
in a subtle manner.

After these first intuitive remarks, let us inspect the RISB results in the cluster regime as summarized
in Fig.~\ref{fig4}. To assess the role of the Hund exchange $J_{\rm H}$ in the given bilayer context, the
two regimes $J_{\rm H}=0$ and $J_{\rm H}=U/4$ are investigated. Because of the challenging numerics,
minimization of the RISB variational problem for interaction strengths as large as handled for solving
the coupled single-site problem proves difficult. However, all the pre-converged calculations for
larger $U$ values provide no obvious signs for physical behavior different from the one discussed below.

In the case of vanishing $J_{\rm H}$
(see Fig.~\ref{fig4}a,b), the additional interlayer correlations from the cluster scheme first leads
to a minor strenghtening of the orbital-correlation differentiation between $d_{z^2}$ and $d_{x^2-y^2}$.
The $d_{z^2}$ orbital becomes stronger occupied and correlated. In general, compared to the finite Hund
exchange cases (cf. Fig.~\ref{fig2}a) the filling of the $\gamma$ band grows stronger for $J_{\rm H}=0$,
but without significant further enhancement in the cluster picture. The local bosonic weights provide
access to the particle-sector multiplet occupations $M_p=\sum_{{\cal A}_p{\cal Q}_p}|\phi_{{\cal A}_p{\cal Q}_p}|^2$
as well as spin-correlation functions. For the former,
nonsurprisingly, the three-particle sector ruling grows with $U$. Notably the two-particle sector,
hosting the $d_{z^2}$ singlet, dominates over the four-particle sector. The local spin moment in the
paramagnetic regime surely grows with $U$, and underlines the enhanced localization character of
$d_{z^2}$ compared to $d_{x^2-y^2}$. The most obvious signature for the tendency to $d_{z^2}$ singlet
formation comes from the enhanced negativity of the interlayer $\langle S_{z^2,\ell=1}S_{z^2,\ell=2}\rangle$
correlation function (see bottom panel of Fig.~\ref{fig4}b). But it appears to start saturating at still
moderate values, not providing clues for strong singlet formation up to a fully-localized regime. Note
that $\langle S_{x^2-y^2,\ell=1}S_{x^2-y^2,\ell=2}\rangle$ remains very small (since here mainly hopping driven) and
$\langle S_{z^2,\ell=1}S_{x^2-y^2,\ell=2}\rangle$ is zero. The latter is obvious from $J_{\rm H}=0$, since there
is signwise no prefered spin correlation between the different orbitals on different sites in the present
case. In the end, the too-weak local-moment formation for vanishing $J_{\rm H}$ in the given multiorbital
setting with the more-itinerant $d_{x^2-y^2}$ electron still favorably scattering with $d_{z^2}$ electrons,
hinders a localized-singlet formation.

For a large part, the $J_{\rm H}=U/4$ results in Fig.~\ref{fig4}c,d mainly differ quantitatively. The
orbital differentiation compared to the single-site results concerning the correlation strength is somewhat
stronger than for vanishing Hund exchange. The tendency for aligning orbital occupations with finite
$J_{\rm H}$ in the single-site regime is counterinteracted on by the interlayer correlations, as displayed
in the mid panel of Fig.~\ref{fig4}c. Filling of the $\gamma$ band increases slighly in the cluster regime,
but still remains far from complete filling, differing with the result of the recent cluster-DMFT
study~\cite{ryee24}. The particle-sector resolved and summed multiplet weight shows minor increased
asymmetries between the 2,4 and 1,5 sectors. Local spin moments are overall increased to some extent,
as expected from the impact of Hund exchange in multiorbital systems. But the values for
$\langle S_m^2\rangle$ are still
far off the fully-localized-spin limit of $s(s+1)=3/4$. The spin correlation function in favor of
$d_{z^2}$ singlet formation is however not strikingly different from the $J_{\rm H}=0$ result. Yet it
is also seen in the bottom panel of Fig.~\ref{fig4}d that the interlayer interorbital spin correlations are
way larger with finite $J_{\rm H}$. As mentioned already in previous works, this is understood from the
fact that the Hund exchange favors aligned spin configurations and the strongly $J_\perp$-driven singlet-like
correlations for $d_{z^2}$ are therefore forced onto $d_{x^2-y^2}$, too.
But importantly, the different actions of $J_{\rm H}$ are finally also detrimental for localized singlet
formation in La$_3$Ni$_2$O$_7$. While it strengthens the {\sl common} local-moment formation, it weakens
a strong orbital polarization which would be needed for a localized $d_{z^2}$ singlet. Moreover, the remaining
$d_{x^2-y^2}$ electron wants to be Hund spin-aligned with either $d_{z^2}$ electron, leading to a frustration
scenario in the localized limit of a $d_{z^2}$ singlet.

The key take away message from the two-site RISB cluster calculations is twofold. First, the sizeable hybridization,
filling scenario and conflicting effects of $J_{\rm H}$ hinders a complete (orbital-selective) $d_{z^2}$ singlet
localization. Second, interlayer correlations under these cirumstances do not have a very strong effect,
rendering the main qualitative results of the standard cluster calculation effectively similar to the ones of
the coupled single-site computation. 

\subsection{Minimal cluster picture\label{res3}}
The two-site cluster approach to electronic correlations in La$_3$Ni$_2$O$_7$ based on the canonical
four-orbital Hamiltonian (\ref{eq:ham1}) provides valuable insights beyond the coupled single-site
picture concerning the interplay of especially the interlayer exchange $J_\perp$ with the
Hund coupling $J_{\rm H}$. 
Yet this standard cluster approach is numerically demanding and schematically not perfectly faithful
to a true low-energy perspective. Even in the liberate Wannier picture, an on-site orbital basis is usually
connected to a high-energy perspective for systems with reasonable intersite correlations. Tailored approaches
which include such intersite correlations within an molecular-orbital setting of the given correlation problem
have therefore already been persued by previous works~\cite{wangkunzhang24,liaowanglei24}. While
insightful, there the correspondingly adapted orbital bases though were mostly chosen/constructed in an
ad-hoc manner.
%%%%%%%%%%%%%%%%%%%%%%%%%%%%%%%%%%%%%%%%%%%%%%%%%%%%%%%%%%%%%%
\begin{figure}[t]
  \includegraphics[width=\linewidth]{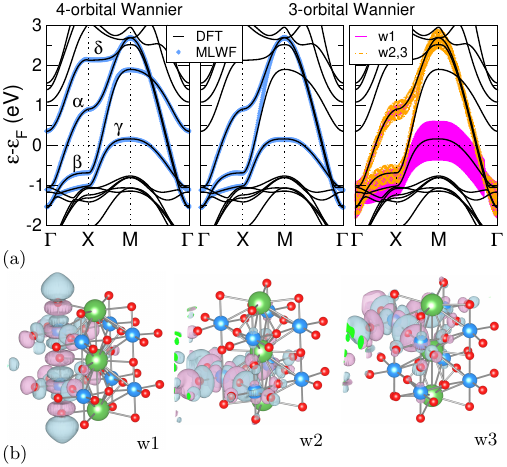}
  \caption{Setting of the MC picture. (a) Wannier-band (in blue) comparison between the canonical four-orbital
    downfolding (left panel) and the MC-relevant three-orbital downfolding (mid panel). Right panel: fatbands
    for the $d_{z^2}$-related $w1$ (magenta) and $d_{x^2-y^2}$-related $w2,3$ (orange) Wannier orbitals.
  (b) Constant$(\pm$)-value surfaces of the Wannier-function basis $\{w1,w2,w3\}$.}\label{fig5}
\end{figure}	
%%%%%%%%%%%%%%%%%%%%%%%%%%%%%%%%%%%%%%%%%%%%%%%%%%%%%%%%%%%%%

In order to work out a minimal cluster (MC) picture with the help of a molecular-orbital basis that carries
interlayer physics by design, we follow here a formal route guided by the concrete low-energy DFT dispersions.
The left panel of Fig.~\ref{fig5}a shows the band structure close to the Fermi level for the
$I4/mmm$ crystal structure of La$_3$Ni$_2$O$_7$ at high pressure~\cite{luhongwang24}, together with the
matching Wannier dispersions based on the four-orbital tight-binding Hamiltonian. Obviously, the Fermi
surface is only formed by the three $\alpha,\beta,\gamma$ bands, and the $\delta$ band remains also unoccupied
with interactions for $n=3$ as discussed in the previous sections.
Remember that the $\delta$ band connects to the ``true'' antibonding part of Ni$(d_{z^2})$-O$(p_z)$-Ni$(d_{z^2})$
(cf. Fig.~\ref{fig1}b,c) with the corresponding bonding part deep down in the occupied part of the spectrum. On
the other hand, the $\gamma$ band is mainly associated with the nonbonding part of
Ni$(d_{z^2})$-O$(p_z)$-Ni$(d_{z^2})$.
Thus, abandoning the $\delta$ band from a true low-energy modeling appears acceptable for
energetical/occupational and symmetry reasons. This can be straightforwardly achieved in the MLWF
construction by guiding a single Wannier centre with dominant $d_{z^2}$ character to the bridging-O site,
instead of two such Wannier centres guided towards Ni$(\ell=1,2)$. The resulting effective three-band Wannier
dispersion on top of the DFT bands is visualized in the mid panel of Fig.~\ref{fig5}a, and the corresponding
Wannier orbitals $w1$-$3$ are depicted in Fig.~\ref{fig5}b. Of course, a minimal three-orbital basis cannot
exactly reproduce the higher-lying DFT bands that dominantly form the previous $\delta$ band, but the
low-energy regime around $\varepsilon_{\rm F}$ is as well  described as in the original four-orbital on-site
basis. The ``interlayer orbital'' $w1$ is the most-true Wannier form of the nonbonding
Ni$(d_{z^2})$-O$(p_z)$-Ni$(d_{z^2})$ molecular orbital. It has dominant weight on the $\gamma$ band (see right
panel of Fig.~\ref{fig5}a), but notably still contributes also to the $\alpha,\beta$ bands which are
majorly formed by $w2,3$. The latter orbitals with dominant $d_{x^2-y^2}$ character and sign-difference
with respect to the nearest-neighbor in-plane O sites are still localized on Ni$(\ell=1,2)$.
These orbitals are degenerate and the crystal-field splitting $\Delta_{\rm CF}=\varepsilon_{w1}-\varepsilon_{w2,w3}$
amounts to $-536$\,meV. This much larger $\Delta_{\rm CF}$ compared to the original four-orbital model is
easily understandable from the fact, that the previous interlayer $d_{z^2}$ hopping is included in the
$w1$ on-site level energy of the MC picture.

Concerning the fully interacting Hamiltonian in this minimal setting, we follow a pragmatic approach
to keep the discussion easily tractable. Importantly, the derived MC model operates at half filling
$n=3$, different from the original $3/8$-filled four-orbital model. Then usually for a cluster
approach (see also previous section), intersite interactions are absent. However it is intuitively clear,
that if we follow that setting also here in the MC picture for $n=3$, the model will quickly run into
a Mott insulator with half-filled $w$ orbitals when increasing $U$.
But this is surely not the physics posed by La$_3$Ni$_2$O$_7$. There, we e.g. expect minor/modest charge
transfers around the $d_{z^2}$ singlet, where the latter translates into a fully-occupied $w1$ orbital.
Therefore, the MC interaction vertex has to be of balanced intra- and interorbital kind. While this
amounts to an explicit intersite interaction e.g. for the $w2,3$ orbitals, it is not at odds with the
low-energy physics. Since for instance, the $w1$ orbital is a ``condensed'' molecular orbital for the
individual $d_{z^2}$, encircling both Ni$(\ell=1,2)$ sites equally. An ``on-site'' $w1$ interaction thus
already carries intersite interactions possibly active in the previous four-orbital model.

Instead of a (numerical) downfolding transformation of the original multi-site
Slater-Kanamori Hamiltonian (\ref{eq:intham}) into the novel $\{w1,w2,w3\}$ basis, which generally
generates intersite and/or additional new interaction terms, we assume a standard
multiorbital Hubbard-model structure. A basic and natural choice is the three-orbital Hubbard model
with density-density interactions, i.e.
\begin{eqnarray}
H^{\rm (MC)}_{\rm int}&=&U\sum_{w} n_{w\uparrow}n_{w\downarrow} \nonumber\\    
&&\hspace*{-0.5cm}+\frac{1}{2}\sum_{w\neq w',\sigma}              
\left\{U' \, n_{w\sigma} n_{w' \bar \sigma}+\, U'' \,n_{w\sigma}n_{w' \sigma}\right\}\;.           
\label{eq:intmc}          
\end{eqnarray} 
It carries the important interorbital interactions, guaranteeing that the MC model does not immediately
run into the half-filled orbital limit. Besides taking care of a matching filling scenario, to be
qualitatively consistent with the two-site four-orbital cluster description, the intersite spin
correlations as displayed in bottom panel of Fig.~\ref{fig4} should also have the same signature. However,
from a standard Hund-rule setting of (\ref{eq:intmc}) in an usual on-site context, all interorbital,
and notably in the MC picture {\sl intersite}, spin correlations will turn out positive. Therefore, we
choose by hand the reversed parametrization of $U'$ and $U''$, namely $U'=U-3J_{\rm mc}$ and $U''=U-2J_{\rm mc}$,
whereby $J_{\rm mc}$ denotes the interorbital(-site) exchange integral in the MC picture. This ensures, that e.g.
$\langle S_{w2}S_{w3}\rangle\sim\langle S_{x^2-y^2,\ell=1}S_{x^2-y^2,\ell=2}\rangle$ is of antiferromagnetic
tendency. Or in other words, albeit employing orthogonal orbitals, the MC model is more aligned with a Heitler-London-like than a Hund-limit framing~\cite{fazekasbook}. The need for this choice renders the relevance of internal frustration physics in La$_3$Ni$_2$O$_7$ obvious. Furthermore, this frustration does not only hold on the intersite level.
It concerns also again the competition between on-site Hund vs. intersite exchange, since one cannot
optimize both simultaneously in the MC picture for $d_{z^2}$ and $d_{x^2-y^2}$.
%%%%%%%%%%%%%%%%%%%%%%%%%%%%%%%%%%%%%%%%%%%%%%%%%%%%%%%%%%%%%%
\begin{figure}[t]
  \includegraphics[width=\linewidth]{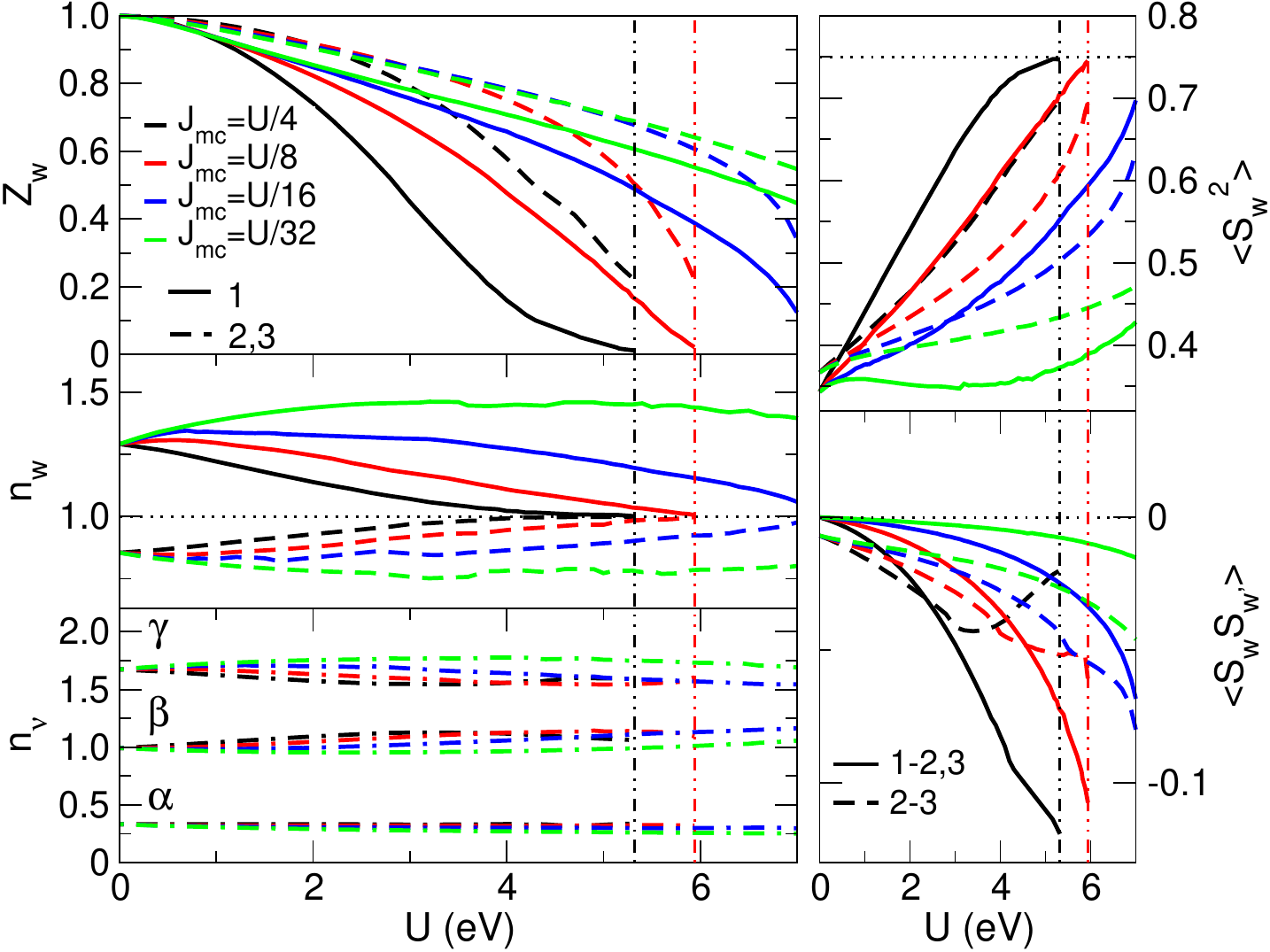}
  \caption{Effect of correlations in the MC model with increasing $U$ and different
    ratios $U/J_{\rm mc}$. Left panel: QP weight $Z_m$ (top) and orbital occupation
        $n_m$ (middle) for $w1$ (full lines) and $w2,3$ (dashed lines),
    as well as QP band occupations $n_\nu$ (bottom) for $\nu=\alpha,\beta,\gamma$.
    Right panel: Orbital-resolved local spin moment
    $\langle S^2_w\rangle$ (top), with line styles as in left panel, and interorbital spin
    correlation $\langle S_wS_{w'}\rangle$. The vertical dashed-double-dottted lines (dark,red)
    refer to ritical $U$, respectively.}\label{fig6}
\end{figure}	
%%%%%%%%%%%%%%%%%%%%%%%%%%%%%%%%%%%%%%%%%%%%%%%%%%%%%%%%%%%%%

Figure~\ref{fig6} displays the results of the RISB calculations for the MC picture, employing different ratios
of $U/J_{\rm mc}$. Since $J_{\rm mc}$ is an effective intersite exchange integral, a value much lower than for the
material-realistic Hund $J_{\rm H}\sim 1$\,eV is expected. Therefore, the operational regime of the MC model
for La$_3$Ni$_2$O$_7$ should be located in the domain of larger $U/J_{\rm mc}$. In line with this, it is observed
that for a comparatively small $U/J_{\rm mc}=4,8$, the system rather quickly runs into locally half-filled
Mottness with increasing $U$. Notably, the appearing Mott transition is of orbital-selective character, i.e.
the $w1$ orbital becomes Mott insulating while $w2,3$ remain metallic. This becomes apparent not only from
the behavior of the QP weight, but also from the local moment $\langle S^2_w\rangle$, which attains the
localized-limit value $3/4$ for orbital $w1$ at the critical $U$. Consequently, the infamous $\gamma$ band
becomes heavily hole doped in that bilayer-unrealistic interaction scenario. While such a half-filled Mott-critical
state is then expected for {\sl any} $U/J_{\rm mc}$ {\sl in the limit of large} $U$, the model behavior for
{\sl large} $U/J_{\rm mc}$ {\sl and intermediate} $U$ may still be fitting with our achieved low-energy understanding
of La$_3$Ni$_2$O$_7$.

And indeed, as observable especially for the case $U/J_{\rm mc}=32$, the occupation of the interlayer $w1$ orbital
remains away from half filling for materials-realistic interaction strenghts. The filling of the corresponding
$\gamma$ band increases from the noninteracting value, but still stays away from complete filling. The correlation
strength also remains moderate, just as before in the coupled single-site and standard-cluster calculations.
Interestingly, the $w1$ local-moment remains rather small for reasonable $U$ in that case, even decreases with
increasing $U$ in some range. Yet this is understandable from the fact that the $w1$ orbital hosts the possible
$d_{z^2}$ singlet with $S=0$, and tendencies towards this singlet formation just reflects the nickelate bilayer
physics. Also the interorbital/site $\langle S_wS_{w'}\rangle$ with stronger negative $w2$-$w3$ are in agreement
with the previous standard-cluster picture. Therefore, the MC picture may transport the essential low-energy
physics of normal-state La$_3$Ni$_2$O$_7$ prone to the superconducting instability. Within RISB, we expect
a $U\sim 4-6$\,eV for this three-orbital low-energy model at nominal half filling, compared to a larger
$U\sim 8-10$\,eV for the canonical four-orbital model at filling $3/8$. This would yield
$J_{\rm mc}\sim 0.1-0.2$\,eV and in a reasonable physical ballpark for such a (quasi-)intersite exchange.
Note that the (critical) interaction strength is usually overestimated within RISB as a saddle-point
method. Smaller such Coulomb integrals are in this respect expected in e.g. DMFT studies of the model.

The MC picture sheds further light on the potential importance of the nonbonding $w1$ band for
unconventional superconductivity. Within the four-band picture this band is often misinterpreted as
stemming from the bonding $d_{z^2}$-orbitals and its presence at the Fermi level appears to be almost
accidental, being subject to a fine tuning of the layer hybridization and the crystal field splitting.
Within the MC approach, the nonbonding character of this band and its strong Ni-O hybridization point towards
its rigidity in the low-energy spectrum. It is likely of potential crucial importance for the occurrence of
high-T$_c$ superconductivity. This is further confirmed by the fact that in the superconducting thin films
under compressive strain, this band appears to be present at the Fermi level~\cite{lipeng25}.  

\section{Conclusions}
A deeper look at the key features and ingredients of the low-energy La$_3$Ni$_2$O$_7$ physics close to the
superconducting state has been presented in this work. From these studies, the nickelate bilayer is characterized
as a rather unique multiorbital and multisite correlation problem. First, the bilayer architecture already
sets the stage for an intriguing competition between on- and intersite correlation physics. Second, the multiorbital
context adds just another level of complexity. And finally, the non-half filling of $3/8$ for the two Ni sites
in the different layers raises sophistication even further. Note that the latter filling is intriguingly
connected to the ligand-hole physics in the compound, i.e. still another (hidden) difficulty. As a result,
we observed manifest frustration behavior from several perspectives. There is a crucial competition between
the interlayer exchange, most-effective for the Ni-$d_{z^2}$ orbitals, and the on-site Hund exchange $J_{\rm H}$.
It effectively leads to the Ni-$d_{z^2}$ electrons remaining itinerant, hindering a full localization of the
$d_{z^2}$ interlayer singlet. Concomitantly, the (flat) $\gamma$ band of nonbonding character remains tied to the
Fermi level for all
investigated model variations. It is important to note in that regard, that the Hund $J_{\rm H}$ is not only
relevant for the sign and size of the spin coupling. In multiorbital Hubbard models it is also instrumental
for the general local-moment formation as well as the degree of orbital polarization. Those two issues
have additionally to be considered carefully when discussing the role of $J_{\rm H}$.

Different to cluster vs. single-site correlation models in cuprates, the straightforward two-site two-orbital
cluster approach, while insightful concerning exchange mechanisms and frustration effects, does not add qualitatively
new physics beyond the coupled single-site approach. Instead, we derived a novel minimal-cluster (MC) model
tailored to the low-energy setting of La$_3$Ni$_2$O$_7$.
It consistes of three effective orbitals, works at half filling and carries the important competition
between on-site and interlayer correlations, while remaining faithful to the original dispersions close to
the Fermi level. Its sole existence already teaches an important lesson about the entangled multiorbital/site
physics of the given system. From there, the (quasi-)intersite exchange between the effective orbitals
attains a peculiar role, and being decisive about the actual correlation regime. The inverted-Hund coupling
of the MC model underlines the inherent frustration physics in the bilayer nickelate. This model was here only
introduced in a semi-intuitive way and studied on its most basic level. Further investigations with more
advanced many-body techniques and/or extension to superconducting instabilities/orders are tasks 
for the future.

\section{Acknowledgements} 
Computations were performed at the Ruhr-University Bochum. F.L. is indebted to C. Piefke for
the original implementation of the combined recursive modified-Broyden+DIIS mixing scheme.

\bibliography{literatur}
\end{document}

%\pagenumbering{Alph}

\title{Supplemental Material: Low-energy perspective of interacting electrons in the normal state of superconducting bilayer nickelate}
\author{Frank Lechermann}
\affiliation{Theoretische Physik III, Ruhr-Universit\"at Bochum,
  D-44780 Bochum, Germany}
\author{Steffen B\"otzel}
  \affiliation{Theoretische Physik III, Ruhr-Universit\"at Bochum,
  D-44780 Bochum, Germany}
\author{Ilya M. Eremin}
\affiliation{Theoretische Physik III, Ruhr-Universit\"at Bochum,
  D-44780 Bochum, Germany}

%\linenumbers  %adds lines to reference changes
\maketitle
\section{Tight-binding description for the minimal-cluster model}

In the following, we provide a minimal three-dimensional tight-binding description of the non-interacting part of the minimal-cluster model. It is based on the introduced Wannier orbitals $w1='1',\, w2='2'$ and $w3='3'$ (see Fig. 5b) and restricted to only a few hoppings. The tight-binding Hamiltonian $H$ has the form
\begin{equation}
H = \sum_\mathbf{k} \Psi_\mathbf{k}^\dagger H_{\mathbf{k}} \Psi_\mathbf{k} = \sum_\mathbf{k} \Psi_\mathbf{k}^\dagger \begin{pmatrix}
T_{11} & V_{12} & V_{23} \\
V_{12} & T_{22} & V_{13} \\
V_{23} & V_{13} & T_{33}
\end{pmatrix}
\Psi_\mathbf{k}
\end{equation}
with $\Psi_\mathbf{k} = (c_{u,\mathbf{k}},c_{z,\mathbf{k}},c_{l,\mathbf{k}})^T$. The matrix $H_{\mathbf{k}}$ is real, symmetric and has components
\begin{align*}
T_{11} &= \varepsilon_{11} + 2t^{(1)}_{11}(\cos k_x  + \cos k_y ) + 4t^{(11)}_{11}\cos k_x \cos k_y  + 2t^{(2)}_{11}(\cos 2k_x  + \cos 2k_y ) \\ 
&+ 4t^{(12)}_{11}(\cos 2k_x \cos k_y  + \cos k_x \cos 2k_y ) 
 + 8t^{\rm (ib)}_{11}\cos \frac{k_x}{2} \cos \frac{k_y}{2} \cos k_{z}  \\
T_{22} &= \varepsilon_{22} + 2t^{(1)}_{22}(\cos k_x  + \cos k_y)  + 2t^{(2)}_{22}(\cos 2k_x  + \cos 2k_y ) + 2t^{(3)}_{22}(\cos 3k_x  + \cos 3k_y ) \\ &+ 2t^{(4)}_{22}(\cos 4k_x  + \cos 4k_y ) 
+ 4t^{(11)}_{22}\cos k_x \cos k_y  + 4t^{(22)}_{22}\cos 2k_x \cos 2k_y  \\
&+ 
4t^{(13)}_{2}\left( \cos k_x \cos 3k_y  + \cos 3k_x \cos k_y \right)  \\
T_{33} &= T_{22} \\
V_{23} &= \varepsilon_{23} + 2t^{(2)}_{23}(\cos 2k_x  + \cos 2k_y ) + 4t^{(11)}_{23}\cos k_x \cos k_y  \\
V_{12} &= 2t^{(1)}_{12}(\cos k_x  - \cos k_y ) + 2t^{(2)}_{12}(\cos 2k_x  - \cos 2k_y ) + 2t^{(3)}_{12}(\cos 3k_x  - \cos 3k_y ) \\
&+ 4t^{(12)}_{12}(\cos 2k_x \cos k_y  - \cos k_x \cos 2k_y ) \\
V_{13} &= -V_{12}.
\end{align*}
The last relation follows because the $w2,3$ Wannier orbitals differ by a negative sign (see Fig. 5b). There is only one sizable interbilayer (ib) hopping, i.e. hopping between different unit cells separated in $c$ direction, which is here denoted as $t^{\rm (ib)}_{11}$. The interbilayer hopping component causes a finite $k_{z}$ dependence. The hopping parameters are given in table \ref{tab:SM1}.

\vspace{5 pt}
\begin{table}[h!]
\centering
\begin{tabular}{|l|p{1cm}|p{1cm}|p{1cm}|p{1cm}|}
\hline
    & \multicolumn{4}{|c|}{orbitals} \\
hopping &\quad \( 11 \) &\quad \( 22 \)  &\quad \( 12 \) &\quad \( 23 \) \\
\hline
\( \varepsilon \) & -160 & 415 &  & -84\\
\( t^{(1)} \) & -120 & -508 & 170 & \\
\( t^{(2)} \) & -4 & -94 & 13 & -7\\
\( t^{(3)} \) & & -17 & 4 & \\
\( t^{(4)} \) & & -6 & &\\
\( t^{(11)} \) & -61 & 135 & & 28 \\
\( t^{(22)} \) & & 10 & &  \\
\( t^{(12)} \) & -4 &  & 5 & \\
\( t^{(13)} \) & & -6 & &\\
\( t^{\rm (ib)} \) & -28 & & & \\
\hline
\end{tabular}
\caption{Hopping parameters in meV. }
\label{tab:SM1}
\end{table}

The Hamiltonian can be diagonalized by using the transformation
\begin{equation}
    U_{\mathbf{k}} = \frac{1}{\sqrt2}
    \begin{pmatrix} 
0 & \sqrt{2} u_- &  - \sqrt{2} u_+ \\
1 & u_+  &  u_- \\
1 & -u_- &  -u_+
    \end{pmatrix}, \quad U_{\mathbf{k}}^\dagger H_{\mathbf{k}}U_{\mathbf{k}} = 
    \begin{pmatrix} 
T_{22}+V_{23} & 0 & 0 \\
0 & \frac{T_{11}+T_{22}-V_{23}}{2} + \Omega_{\mathbf{k}}  & 0 \\
0 & 0 & \frac{T_{11}+T_{22}-V_{23}}{2} - \Omega_{\mathbf{k}}
    \end{pmatrix}
\end{equation}
where
\begin{align}
    u_{\pm} = \sqrt{\frac{1}{2} \left( 1\pm\frac{T_{22}-V_{23}-T_{11}}{2\Omega_{\mathbf{k}}} \right)} \quad , \qquad
    \Omega_{\mathbf{k}} = \sqrt{ \left( \frac{T_{22}-V_{23}-T_{11}}{2} \right)^2+2V_{12}^2} \quad.
\end{align} 

%%%%%%%%%%%%%%%%%%%%%%%%%%%%%%%%%%%%%%%%%%%%%%%%%%%%%%%%%%%%%%
\begin{figure*}[t]
      \includegraphics[width=\linewidth]{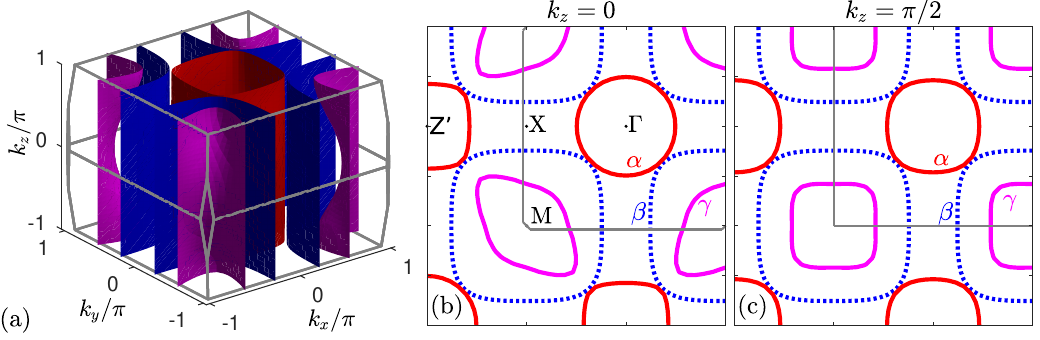}
      \caption{(a) The Fermi surface consists of three sheets. The grey lines indicate the first Brillouin zone. (b),(c) Cuts through the Fermi surface at constant $k_z = 0$ and $k_z = \pi/2$. The boundary of the first Brillouin zone is again indicated by grey lines. The bands are shown using the same colors as in (a).
}\label{figSM1}
\end{figure*}	
%%%%%%%%%%%%%%%%%%%%%%%%%%%%%%%%%%%%%%%%%%%%%%%%%%%%%%%%%%%%%
\begin{figure*}[t]
      \includegraphics[width=\linewidth]{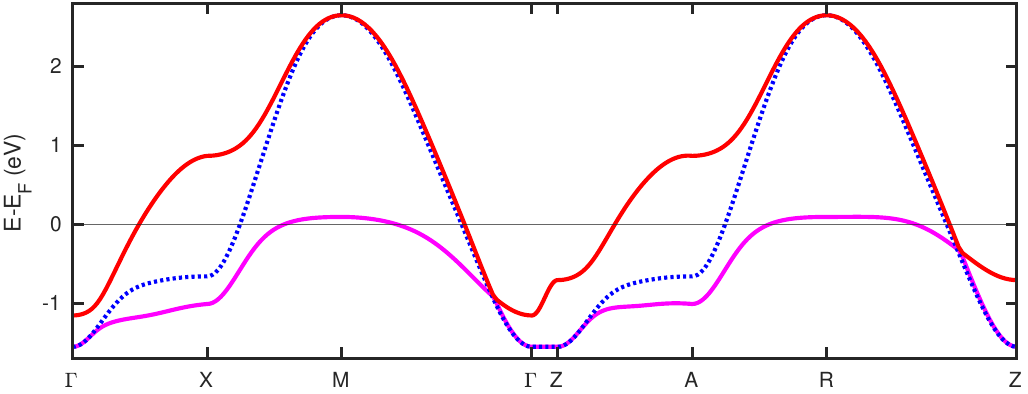}
      \caption{Band structure along high-symmetry path, given by $\Gamma = (0,0,0)$, X $=(\pi,0,0)$, M $=(\pi,\pi,0)$, Z $=(0,0,\pi)$, A $=(\pi,0,\pi)$ and R $=(\pi,\pi,\pi)$.
}\label{figSM2}
\end{figure*}	
%%%%%%%%%%%%%%%%%%%%%%%%%%%%%%%%%%%%%%%%%%%%%%%%%%%%%%%%%%%%%

The Fermi surface is shown in Fig.~\ref{figSM1} and the band structure along a high symmetry line in Fig.~\ref{figSM2}. The $w_1$ dominated $\gamma$ sheet, which is shown in magenta in Fig.~\ref{figSM1}, has the largest $k_{z}$ dependence of all Fermi surface sheets. The $\alpha$ pocket is shown in red and also obeys finite $k_{z}$ dependence. This can be seen in the $k_z = 0$ Fermi surface cut shown in Fig.~\ref{figSM1}(b) by comparing the $\alpha$ pocket in the first Brillouin zone around the $\Gamma$ pocket and the $\alpha$ pocket in the neighboring zones centered around Z'. The $\beta$ pocket shown in blue does not show any $k_z$ dependence. In Fig.~\ref{figSM1}(c) a Fermi surface cut for $k_z = \pi/2$ is shown. At $k_z = \pi/2$, the in-plane momenta obey $\pi$ periodicity. If one restricts to a two-dimensional description, using $k_z = \pi/2$ is most natural because it is the interpolation between the extrema $k_z = 0$ and $k_z = \pi$.  

\bibliography{literatur}